\documentclass[a4paper, 12pt]{article}
\usepackage[cp1251]{inputenc} 
\usepackage[english, russian]{babel} 
\usepackage{amsmath, latexsym,  amssymb, bm, array, graphics, eucal,
amsfonts, amsthm}
\usepackage[flushleft,small]{caption2}

\usepackage[dvips]{graphicx}
\makeatletter
\newcommand{\rot}{\mathop{\rm rot\, }}
\newcommand{\Res}{\mathop{\rm Res\,}}
\newcommand{\const}{\mathop{\rm const\, }}

\renewcommand{\Re}{\mathop{\rm Re\,}}
\renewcommand{\Im}{\mathop{\rm Im\,}}

\usepackage[dvips]{graphicx}
\makeatother

\begin{document}
\thispagestyle{empty} \large
\newcommand{\mc}[1]{\mathcal{#1}}
\newcommand{\E}{\mc{E}}

\begin{center}
{\Large \textbf{Behaviour of plasma with arbitrary degree of degeneration of electronic gas
in a slab of conducting medium}}
\medskip

{\large \textbf{A. V. Latyshev$^*$, N. M. Gordeeva$^{**}$}}
\medskip

\textit{$^*$ Moscow State Regional University
\\$^{**}$ Bauman Moscow State Technical University}

\end{center}

\thispagestyle{empty}
\large
\renewcommand{\refname}{\begin{center} \center \bf  REFERENCES\end{center}}

The analytical solution of the boundary problem on behaviour (oscillations) of the
electronic plasmas with arbitrary degree of degeneration of electronic gas in a slab
of the conducting medium is received.
The kinetic Vlasov---Boltzmann equation with integral of collisions
type BGK (Bhatnagar, Gross and Krook) and  Maxwell equation  for electric field are applied .
Mirror (reflections) boundary conditions are used.


\begin{center}
  {\bf 1. Introduction}
\end{center}

Целью настоящей работы является аналитическое решение граничной задачи о поведении
невырожденной электронной плазмы Ферми---Дирака в слое проводящей среды. Слой находится
во внешнем переменном электрическом поле. Рассматривается случай зеркального
отражения электронов от границы плазмы.
Проведен анализ существования дискретных мод задачи.
Оказалось, что, во-первых, мода Друде существует при всех значениях параметров задачи.
Во-вторых, выяснено, что допустимые значения параметров задачи разделены кривой на две
области, в одной  из которыхмода Дебая существует, а в другой отсутствует.

Разложение решения граничной задачи по собственным функциям представляет собой сумму
линейной комбинации дискретных (частных) решений, отвечающих дискретному спектру  или
спектру, присоединенному к непрерывному, и интеграла по непрерывному спектру.
Отыскание коэффициентов дискретного и непрерывного спектров с помощью граничных условий
означает построение функции распределения и электрического поля в явном виде.

В работе выяснена структура экранированного электрического поля и функции
распределения электронов.

Задача о колебаниях электронной плазмы впервые рассматривалась А.А. Власовым
\cite{Vlasov}. В \cite{Landau} Л.Д. Ландау аналитически решил задачу о поведении
бесстолкновительной плазмы в полупространстве, находящемся во внешнем продольном
(перпендикулярно поверхности) электрическом поле, и при зеркальном отражении
электронов от границы. Поэтому задачу о колебаниях плазмы естественно называть
задачей Власова---Ландау.

Впервые задача о колебаниях вырожденной электронной плазмы в слое металла была аналитически
решена в \cite{LTMF1992}. Так что настоящая работа является продолжением работы
\cite{LTMF1992} и обобщением ее результатов на общий случай плазмы с произвольной
степенью вырождения электронного газа.

С диффузным граничным условием для столкновительной плазмы эта задача в
полупространстве была рассмотрена впервые в \cite{Poverhnost93}.
В \cite{Poverhnost93} рассматривались общие вопросы разрешимости данной задачи с диффузными
граничными условиями, исследована структура дискретного спектра в зависимости от
параметров задачи. Детальный анализ решения в общем случае в указанной работе не
проводился ввиду сложного характера этого решения.

Полное решение задачи для вырожденной и максвелловской плазмы дано соответственно в
работах \cite{LTMF2006} и \cite{MGG2006}.
Наиболее полное изложение аналитического решения полупространственной граничной
задачи  о колебаниях электронного газа представлено  в \cite{Enziklopedia}.

Родственные задачи возникают в вопросах отражения
электромагнитного поля и были рассмотрены в работах \cite{Kliewer1975}, \cite{Kliewer1970}
методом интегральных преобразований.

В работах \cite{Gohfeld1}, \cite{Gohfeld2} был проведен общий асимптотический анализ поведения
электрического поля на большом расстоянии от поверхности. В \cite{Gohfeld1} указано на
особое значение анализа поведения поля вблизи плазменного резонанса. При этом в
работе \cite{Gohfeld2} выяснено, что поведение поля в этом случае для случаев
зеркального и диффузного рассеяния электронов на поверхности существенно отличается.

Данная задача имеет большое значение в теории плазмы (см., например, \cite{Abrikosov},
\cite{Kadomtsev}) и продолжает изучаться в различных постановках и в настоящее время
\cite{Chizhonkov}.

\begin{center}
\bf 2. Statement of the problem and the basic equations
\end{center}

Пусть невырожденная плазма Ферми---Дирака занимает слой $|x|<L$, заполненный проводящей
средой. Будем считать внешнее поле достаточно слабым, чтобы было применимо
линейное приближение \cite{Abrikosov}.
Будем использовать $\tau$--модельное уравнение Власова---Больцмана:
$$
\dfrac{\partial f}{\partial t}+ {\bf v} \dfrac{\partial f}{\partial {\bf r}}+
e\Big({\bf E}+\dfrac{1}{c}[{\bf v,H}]\Big)\dfrac{\partial f}{\partial
{\bf p}}=\nu\Big(f_{eq}-f\Big),
\eqno{(2.1)}
$$
и уравнение Максвелла для электрического поля
$$
{\rm div} \;{\bf E}=4\pi \rho,\quad
\rho=e\int(f-f_0)d\Omega_F,\qquad d\Omega_F=\dfrac{(2s+1)d^3p}{(2\pi \hbar)^3}.
\eqno{(2.2)}
$$

Здесь $f_{eq}$ -- локально--равновесная функция распределения Ферми---Дирака,
$$
f_{eq}(x,v,t)=\left\{1+\exp\dfrac{\E-\mu(x,t)}{kT}\right\}^{-1},
$$
$f_0=f_{FD}$ -- невозмущенная (абсолютная) функция распределения Ферми---Дирака,
$$
f_0(v,\mu)=f_{FD}(v,\mu)=\left\{1+\exp\dfrac{\E-\mu}{kT}\right\}^{-1},
$$
${\bf p}=m{\bf v}$ -- импульс электрона, $\E={mv^2}/{2}$ -- кинетическая энергия
электрона, $\mu$ и $\mu(x,t)$ -- соответственно невозмущенный и возмущенный
химический потенциал, $e$ и $m$ -- заряд и масса электрона, $\rho$ -- плотность
заряда, $\hbar$ -- постоянная Планка, $\nu$ -- эффективная частота рассеяния
электронов, $s$ -- спин частиц, для электрона $s=1/2$, $k$ -- постоянная Больцмана,
$T$ -- температура плазмы, которая считается постоянной в данной задаче, ${\bf
E}(x,t)$ и ${\bf H}(x,t)$ -- электрическое и магнитное поля внутри плазмы.

Внешнее электрическое поле вне плазмы перпендикулярно границе плазмы и меняется по
закону $ {\bf E}_{ext}(t)=E_0e^{-i\omega t}(1,0,0)$. Соответствующее
самосогласованное электрическое поле внутри плазмы будем обозначать через $ {\bf
E}(x,t)=E(x)e^{-i\omega t}(1,0,0).$ Нетрудно проверить, что при выбранной конфигурации
внешнего электрического поля поля
${\bf H}=-(ic/\omega)\rot {\bf E}=0$. Таким образом, магнитное поле
не входит в уравнение (2.1).

Так как внешнее поле имеет одну $x$--компоненту, то функция распределения $f$ имеет
вид $f=f(x,v_x,t)$, $v_x$ -- проекция скорости электронов на ось $x$. Внешнее
электрическое поле вызывает изменение химического потенциала
$
\mu(x,t)=\mu+\delta \mu(x)e^{-i\omega t}, \; \mu=\const $ -- значение химического
потенциала, отвечающее отсутствию внешнего электрического поля на границе плазмы.
Для приведенного химического потенциала предыдущее равенство имеет вид
$
\alpha(x,t)=\alpha+\delta \alpha(x)e^{-i\omega t},\; \alpha=\const.
$
Будем считать, что величина $\delta \alpha(x,t)=\delta \alpha(x)e^{-i\omega t}$
--- возмущение приведенного химического потенциала является малым параметром, т.е.
$|\delta \alpha(x,t)|=|\delta \alpha(x)|\ll 1$. Физически это неравенство означает,
что возмущение химического потенциала много меньше тепловой энергии электронов:
$
|\delta\mu(x,t)|\ll \E_T, \; \E_T={mv_T^2}/{2}.
$
 Будем действовать методом последовательных приближений, считая, что $|\delta\alpha
(x)|\ll 1$.

Линеаризацию уравнений (2.1) и (2.2) проведем относительно
абсолютной функции распределения Ферми---Дирака $f_0$. 
Введем безразмерный импульс (скорость) электронов ${\bf P}={\bf p}/{p_T}$ = ${\bf
v}/{v_T}$, $v_T$ -- тепловая скорость электронов, $v_T=\sqrt{{2kT}/{m}}$, и
безразмерный (приведенный) химический потенциал $\alpha={\mu}/{kT}$. Линеаризуем
локально-равновесную функцию распределения:
$
f_{eq}(x,P,t)=f_0(P,\alpha)+g(P,\alpha)\delta \alpha(x)e^{-i\omega t},
$
где
$$
f_0(P,\alpha)=f_{FD}(P,\alpha)=\dfrac{1}{1+e^{P^2-\alpha}}, \qquad
g(P,\alpha)=\dfrac{e^{P^2-\alpha}}{(1+e^{P^2-\alpha})^2}.
$$

Линеаризуем функцию распределения электронов:
$$
f(x,P_x,t)=f_0(P,\alpha)+g(P,\alpha)h(x,P_x)e^{-i\omega t},
\eqno{(2.3)}
$$
где $h(x,P_x)$ -- новая неизвестная функция.
Заметим, что в линейном приближении
$$
e{\bf E}\dfrac{\partial f}{\partial {\bf p}}=
e{\bf E}\dfrac{\partial f_0}{\partial {\bf p}}=
\dfrac{e}{p_T}{\bf E}\dfrac{\partial f_0}{\partial {\bf P}}=
-E(x)e^{-i\omega t}\dfrac{2e P_x}{p_T}g(P,\alpha).
$$

Вместо уравнений (2.1) и (2.2) с помощью (2.3) и предыдущих соотношений получаем:
$$
-i\omega h(x,P_x)+v_TP_x\dfrac{\partial h}{\partial x}+\nu h(x,P_x)=
e E(x)\dfrac{2P_x}{p_T}+\nu \delta \alpha(x),
\eqno{(2.4)}
$$

$$
\dfrac{d E(x)}{d x}=\dfrac{8\pi ep_T^3}{(2\pi\hbar)^3}
\int h(x,P_x)g(P,\alpha)\,d^3P.
\eqno{(2.5)}
$$

Величина $\delta \alpha(x)$ определяется из закона сохранения числа частиц:
$$
\int f_{eq}\,d\Omega_F=\int f\,d\Omega_F.
$$

Из этого уравнения находим, что
$$
\delta \alpha(x)=\int h(x,P_x)g(P,\alpha)d\Omega_F
\Big[\int g(P,\alpha)d\Omega_F\Big]^{-1}.
$$

Здесь
$$
\int g(P,\alpha)d^3P=4\pi g_2(\alpha),\quad g_2(\alpha)=\int_{0}^{\infty}g(P,\alpha)P^2dP=
\dfrac{1}{2}s_0(\alpha),
$$
где
$$
s_0(\alpha)=\int_{0}^{\infty}\dfrac{dP}{1+e^{P^2-\alpha}}=
\int_{0}^{\infty} f_0(P,\alpha)dP.
$$

Отсюда ясно, что
$$
\delta \alpha(x)=\dfrac{1}{4\pi g_2(\alpha)} \int h(x,P_x)g(P,\alpha)d^3P.
$$

Вычисляя здесь внутренний двойной интеграл в плоскости $(P_y,P_z)$, получаем, что
$$
\delta \alpha(x)=\dfrac{1}{2s_0(\alpha)}\int_{-\infty}^{\infty}f_0(P_x,\alpha)
h(x,P_x)dP_x.
$$

Далее будем полагать, что $E(x)=E_0e(x)$.
Система уравнений (2.4) и (2.5) преобразуется к следующему виду:
$$
v_TP_x\dfrac{\partial h}{\partial x}+(\nu-i\omega)h(x,P_x)=\dfrac{2eE_0}{p_T}P_xe(x)+
$$
$$
+\dfrac{\nu}{2s_0(\alpha)}\int_{-\infty}^{\infty}f_0(P_x,\alpha)h(x,P_x)dP_x,
\eqno{(2.6)}
$$

$$
E_0\dfrac{d e(x)}{d x}=\dfrac{8\pi^2ep_T^3}{(2\pi\hbar)^3}\int_{-\infty}^{\infty}
f_0(P_x,\alpha)h(x,P_x)dP_x.
\eqno{(2.7)}
$$

В уравнениях (2.6) и (2.7) перейдем к безразмерным величинам и функциям:
$$
x_1=\dfrac{x}{\lambda},\quad \lambda=\tau v_T, \quad P_x=\mu,\quad
H(x_1,\mu)=\dfrac{\nu p_T}{2eE_0}h(x,\mu),
$$
где $\lambda$ -- средняя длина свободного пробега электронов (между двумя
последовательными столкновениями).
В результате перехода к безразмерным параметрам и функциям получаем следующую
систему уравнений:
$$
\mu\dfrac{\partial H}{\partial x_1}+w_0H(x,\mu)=\mu e(x_1)+
\int_{-\infty}^{\infty}k(\mu',\alpha)H(x,\mu')d\mu',
\eqno{(2.8)}
$$
$$
\dfrac{d e(x_1)}{d x_1}=
\varkappa^2(\alpha)\int_{-\infty}^{\infty}k(\mu',\alpha)H(x_1,\mu')d\mu'.
\eqno{(2.9)}
$$

В (2.8) и (2.9)
$$
\varkappa^2(\alpha)=\dfrac{32\pi^2e^2p_T^3s_0(\alpha)}{(2\pi\hbar)^3m\nu^2},
$$
введена новая функция $ k(\mu,\alpha)=f_0(\mu,\alpha)/(2s_0(\alpha)), $ обладающая
свойством $ \int_{-\infty}^{\infty}k(\mu,\alpha)d\mu=1, $ кроме того, $w_0=1-i
\omega/\nu=1-i \omega \tau=1-i\Omega/\varepsilon$, где $\Omega=\omega/\omega_p$,
$\varepsilon=\nu/\omega_p$, $\omega_p$ -- плазменная (ленгмюровская) частота,
$
\omega_p=\sqrt{4\pi e^2 N/m}.
$
Здесь $N$ -- числовая плотность (концентрация) электронов в равновесном состоянии.
Поведение ядер уравнений (2.8) и (2.9) при различных значениях химического
потенциала приведено на рис. 1.
\begin{figure}[h]
\center
\includegraphics[width=9cm, height=6.75cm]{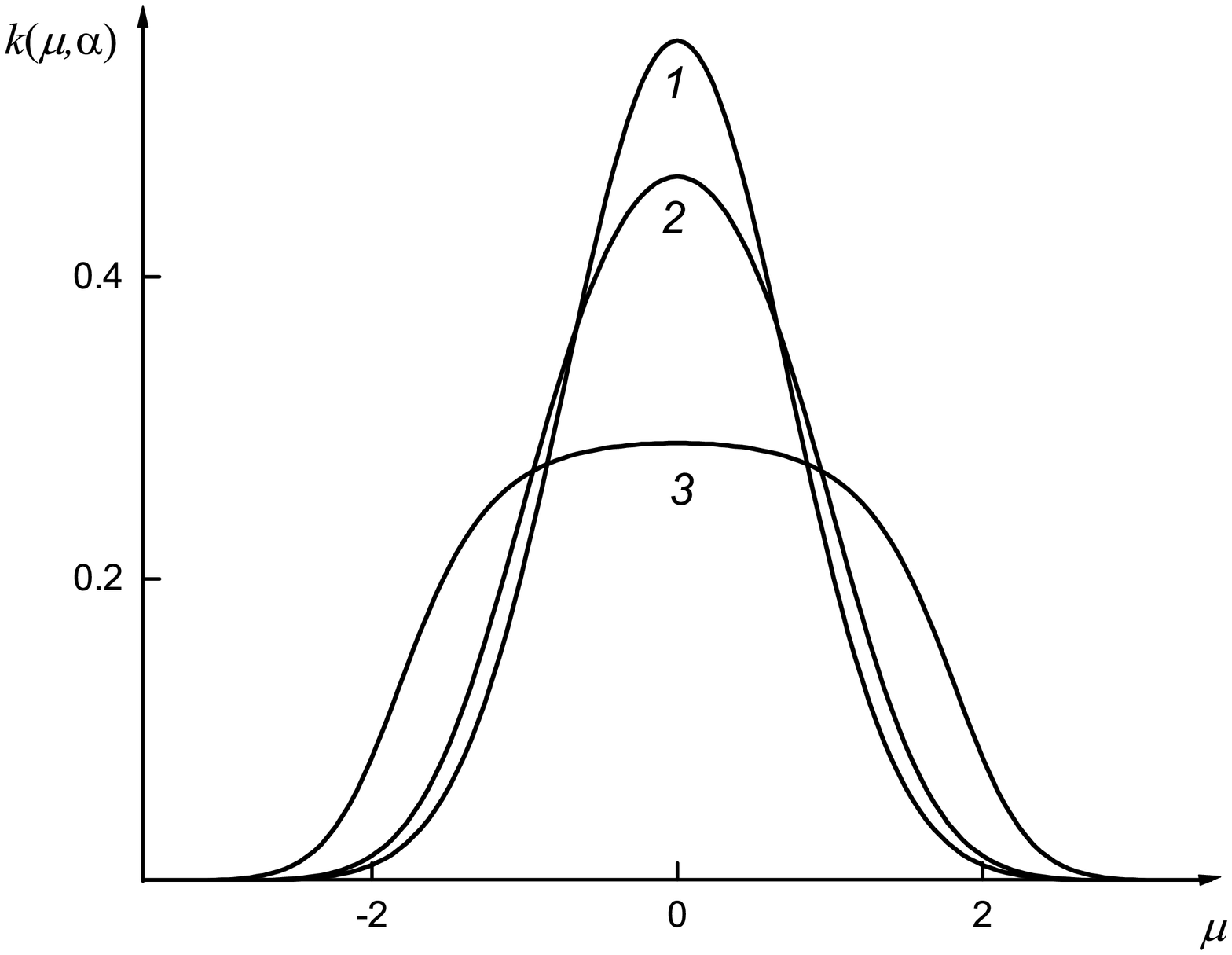}
\center{Рис. 1. Графики функции $k(x,\alpha)$ при $\alpha=-3,0,3$ (кривые 1,2,3)}.
\end{figure}
Из определения числовой плотности вытекает, что
$$
N=\int f_0(P,\alpha)d\Omega_F=
\dfrac{2p_T^3}{(2\pi\hbar)^3}\int\dfrac{d^3P}{1+e^{P^2-\alpha}}=
\dfrac{8\pi p_T^3}{(2\pi\hbar)^3}s_2(\alpha),
$$
где
$$
s_2(\alpha)=\int_{0}^{\infty}\dfrac{P^2dP}{1+e^{P^2-\alpha}}=
\int_{0}^{\infty}P^2 f_0(P,\alpha)dP.
$$

Следовательно, числовая плотность частиц плазмы и тепловое волновое число
$k_T=mv_T/\hbar$ связаны соотношением
$
N=(s_2(\alpha)/\pi^2)k_T^3,
$
кроме того,
$$
\varkappa^2(\alpha)=\dfrac{\omega_p^2}{\nu^2}\dfrac{s_0(\alpha)}{s_2(\alpha)}=
\dfrac{\Omega_p^2}{r(\alpha)}=\dfrac{1}{\varepsilon^2r(\alpha)},
$$
где
$$
r(\alpha)=\dfrac{s_2(\alpha)}{s_0(\alpha)},\qquad
\varepsilon=\dfrac{\nu}{\omega_p}=\dfrac{1}{\tau \omega_p}=\dfrac{1}{\Omega_p},\qquad
\Omega_p=\omega_p\tau.
$$
Графики функций $s_0(\alpha), s_2(\alpha)$ и $r(\alpha)$ изображены на рис. 2.~
\begin{figure}[h]
\center
\includegraphics[width=10cm, height=8cm]{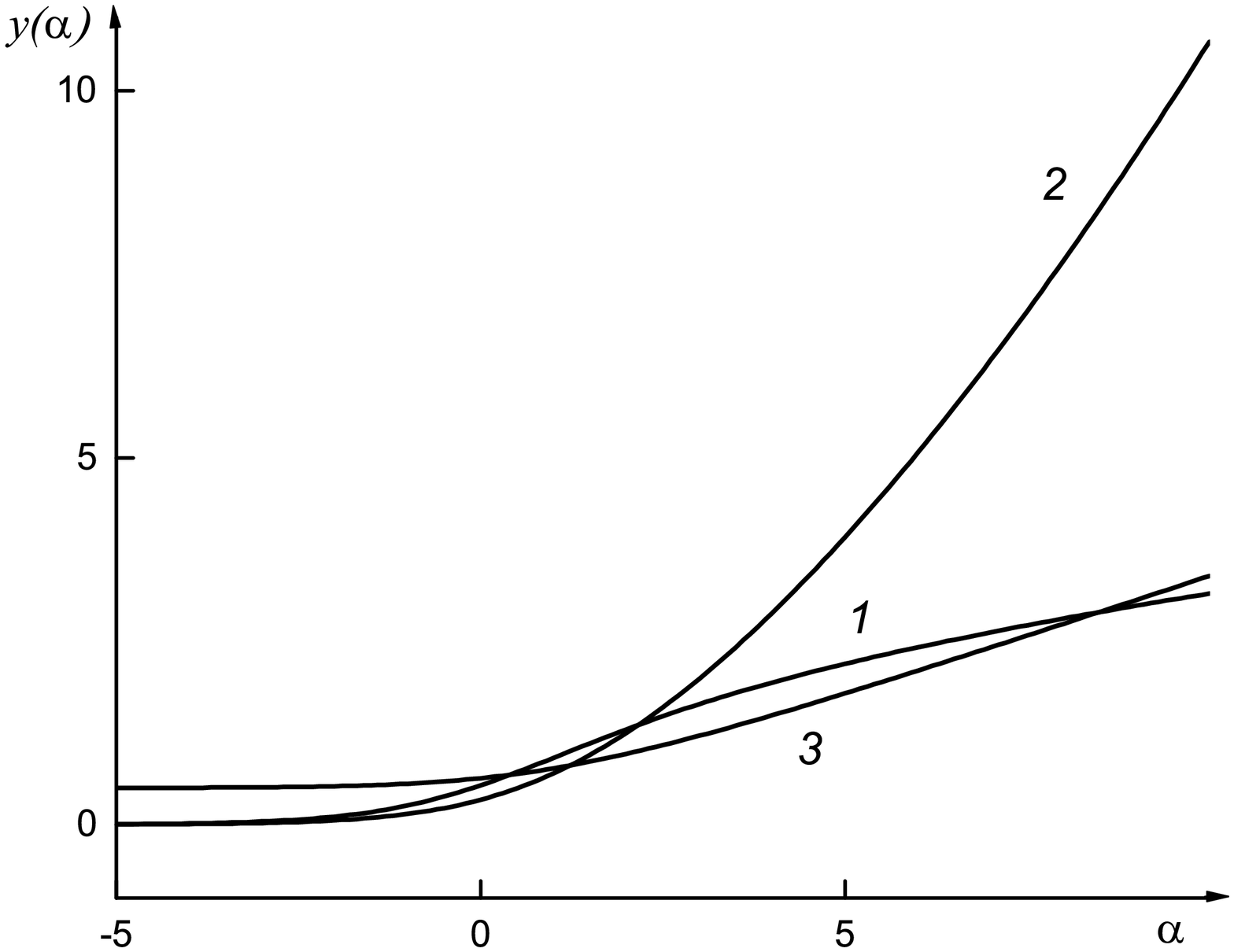}
\center{Рис. 2. Графики функций $s_0(\alpha)$ (кривая 1), $s_2(\alpha)$ (кривая 2)
и $r(\alpha)$ (кривая 3) при изменении приведенного
химического потенциала от $\alpha=-5$ до $\alpha=10$}.
\end{figure}

С помощью этих выражений уравнение (2.9) запишем в виде
$$
\dfrac{d e(x_1)}{d x_1}=\dfrac{1}{\varepsilon^2r(\alpha)}
\int_{-\infty}^{\infty}k(\mu',\alpha)H(x_1,\mu')d\mu'.
\eqno{(2.10)}
$$

Известно, что частота плазменных колебаний как правило много больше частоты
столкновений электронов в металле \cite{Landau10}. Поэтому в случае, когда
$\omega\sim\omega_p$ выполняется условие $\omega_p\gg \nu$. Наиболее типичные
значения величины $\varepsilon$ заключены в интервале: $10^{-4}\leqslant \varepsilon
\leqslant 10^{-2}$.

В случае зеркального отражения электронов от границы плазмы для  функции распределения
электронов имеем следующие граничные условия на границе слоя размера $2L$:
$$
f(\pm L,v_x,v_y,v_z,t)=f(\pm L,-v_x,v_y,v_z,t),\qquad -\infty<v_x<+\infty.
$$

Отсюда для функции $H(x_1,\mu)$ получаем зеркальные граничные условия:
$$
H(l,\mu)=H(l,-\mu), \qquad H(-l,\mu)=H(-l,-\mu),\qquad \mu>0.
\eqno{(2.11)}
$$
Здесь $l=L/\lambda$ -- величина слоя в единицах свободного пробега электронов.

Для электрического поля граничное условие имеет вид:
$$
e(l)=1, \qquad e(-l)=1.
\eqno{(2.12)}
$$

Таким образом, граничная задача о колебаниях плазмы в слое проводящей среды
сформулирована полностью и состоит в нахождении такого  решения уравнений
(2.8) и (2.10), которое удовлетворяет граничным условиям (2.11) и (2.12).\medskip

\begin{center}
{\bf 3. Eigen solutions of the continuous spectrum}
\end{center}\medskip

Сначала будем искать общее решение системы уравнений (2.8) и (2.10).
Разделение переменных согласно общему методу Фурье приводит к следующей подстановке:
$$
H_\eta(x_1,\mu)=\exp\Big(-\dfrac{w_0x_1}{\eta}\Big)\Phi_1(\eta,\mu)+
\exp\Big(\dfrac{w_0x_1}{\eta}\Big)\Phi_2(\eta,\mu),
\eqno{(3.1)}
$$
$$
e_\eta(x_1)=\Bigg[\exp\Big(-\dfrac{w_0x_1}{\eta}\Big)+
\exp\Big(\dfrac{w_0x_1}{\eta}\Big)\Bigg]E(\eta),
\eqno{(3.2)}
$$
где $\eta$ -- спектральный параметр, или параметр разделения, вообще говоря, комплексный.

Подставим равенства (3.1) и (3.2) в уравнения (2.8) и (2.10).
Получаем характеристическую систему уравнений

$$
(\eta-\mu)\Phi_1(\eta,\mu)=
\eta\mu \dfrac{E(\eta)}{w_0}+\dfrac{\eta}{w_0}\int_{-\infty}^{\infty}
k(\mu',\alpha)\Phi_1(\eta,\mu')\,d\mu',
\eqno{(3.3)}
$$

$$
(\eta+\mu)\Phi_2(\eta,\mu)=
\eta\mu \dfrac{E(\eta)}{w_0}+\dfrac{\eta}{w_0}\int_{-\infty}^{\infty}
k(\mu',\alpha)\Phi_2(\eta,\mu')\,d\mu',
\eqno{(3.4)}
$$

$$
-\dfrac{w_0}{\eta}E(\eta)=\dfrac{1}{\varepsilon^2r(\alpha)}\cdot
\int_{-\infty}^\infty k(\mu',\alpha)\Phi_1(\eta,\mu')\,d\mu',
\eqno{(3.5)}
$$

$$
\dfrac{w_0}{\eta}E(\eta)=\dfrac{1}{\varepsilon^2r(\alpha)}\cdot
\int_{-\infty}^\infty k(\mu',\alpha)\Phi_2(\eta,\mu')\,d\mu'.
\eqno{(3.6)}
$$

С помощью (3.5) и (3.6) преобразуем уравнения (3.3) и (3.4). Получаем следующую систему
уравнений:
$$
(\eta-\mu)\Phi_1(\eta,\mu)=\dfrac{E(\eta)}{w_0}(\mu\eta-\eta_1^2),
\eqno{(3.7)}
$$

$$
(\eta+\mu)\Phi_2(\eta,\mu)=\dfrac{E(\eta)}{w_0}(\mu\eta+\eta_1^2).
\eqno{(3.8)}
$$

Здесь

$$
\eta_1^2=w_0\varepsilon^2r(\alpha)=
\dfrac{\nu^2}{\omega_p^2}\Big(1-i\dfrac{\omega}{\nu}\Big)r(\alpha)=\varepsilon
(\varepsilon-i\Omega)r(\alpha).
$$

При $\eta\in (-\infty,+\infty)$ ищем решение характеристических уравнений (3.7) и (3.8)
в пространстве обобщенных функций \cite{Zharinov}:
$$
\Phi_1(\eta,\mu)=\dfrac{E(\eta)}{w_0}(\mu\eta-\eta_1^2)P\dfrac{1}{\eta-\mu}+
g_1(\eta)\delta(\eta-\mu),
\eqno{(3.9)}
$$

$$
\Phi_2(\eta,\mu)=\dfrac{E(\eta)}{w_0}(\mu\eta+\eta_1^2)P\dfrac{1}{\eta+\mu}+
g_2(\eta)\delta(\eta+\mu).
\eqno{(3.10)}
$$

Здесь $\eta,\mu\in (-\infty,+\infty)$. Множество значений $\eta$, заполняющих числовую
прямую, называют непрерывным спектром характеристического уравнения.

В (3.9) и (3.10) $\delta(x)$ -- дельта--функция Дирака, символ $Px^{-1}$ -- означает
главное значение интеграла при интегрировании выражения $x^{-1}$, функции $g_1(\eta)$ и
$g_1(\eta)$  играют роль  произвольной "постоянной"\, интегрирования.

Решения (3.9) и (3.10) уравнения (3.4) называются собственными функциями характеристического
уравнения.

Для нахождения функций $g_1(\eta)$ и $g_2(\eta)$ подставим (3.9) и (3.10) соответственно
в уравнения (3.5) и (3.6). В результате получаем, что
$$
g_1(\eta)=-g_2(\eta),
\qquad
g_2(\eta)=\eta_1^2E(\eta)\dfrac{\Lambda(\eta)}{\eta k(\eta,\alpha)}.
$$

В этих равенствах введена дисперсионная функция
$$
\Lambda(z)=\Lambda(z,\Omega,\varepsilon)=1+\dfrac{z}{w_0\eta_1^2}
\int_{-\infty}^{\infty}\dfrac{\eta_1^2-\mu'z}{\mu'-z}k(\mu',\alpha)d\mu'.
\eqno{(3.11)}
$$

С помощью найденных функций $g_1(\eta)$ и $g_2(\eta)$
собственные функции (3.9) и (3.10) характеристической системы уравнений (3.7) и (3.8)
представим в виде
$$
\Phi_1(\eta,\mu)=\dfrac{E(\eta)}{w_0}F_1(\eta,\mu),\qquad
\Phi_2(\eta,\mu)=\dfrac{E(\eta)}{w_0}F_2(\eta,\mu).
$$

В этих равенствах
$$
F_1(\eta,\mu)=P\dfrac{\mu\eta-\eta_1^2}{\eta-\mu}-
\dfrac{w_0\eta_1^2\Lambda(\eta)}{\eta k(\eta,\alpha)}\delta(\eta-\mu)
$$
и
$$
F_2(\eta,\mu)=P\dfrac{\mu\eta+\eta_1^2}{\eta+\mu}+
\dfrac{w_0\eta_1^2\Lambda(\eta)}{\eta k(\eta,\alpha)}\delta(\eta+\mu).
$$
Семейство (3.1) и (3.2)  собственных решений уравнений (2.8) и (2.10)  отвечает,
как уже указывалось, непрерывному спектру. Это семейство непрерывного спектра часто
называют модой Ван Кампена.

Заметим, что
$F_2(\eta,-\mu)=-F_1(\eta,\mu)$ и $F_2(-\eta,\mu)=F_1(\eta,\mu)$.

Дисперсионную функцию  $\Lambda(z)$ можно представить в виде:
$$
\Lambda(z)=1-\dfrac{1}{w_0}+\dfrac{\eta_1^2-z^2}{w_0\eta_1^2}\lambda_0(z,\alpha),
\eqno{(3.12)}
$$
где
$$
\lambda_0(z,\alpha)=1+z\int\limits_{-\infty}^{\infty}\dfrac{k(\mu,\alpha)d\mu}{\mu-z}.
$$

По формулам Сохоцкого вычислим граничные значения дисперсионной функции сверху и снизу
на действительной оси:
$$
\Lambda^{\pm}(\mu)=\Lambda(\mu)\pm i\pi\dfrac{\mu k(\mu,\alpha)}
{w_0\eta_1^2}(\eta_1^2-\mu^2).
$$
Здесь
$$
\Lambda(\mu)=1+\dfrac{z}{w_0\eta_1^2}
\int\limits_{-\infty}^{\infty}\dfrac{\eta_1^2-\mu'\mu}{\mu'-\mu}k(\mu',\alpha)d\mu',
$$
причем интеграл в этом выражении понимается как особый в смысле главного значения по
Коши.

\begin{center}
\bf 4. Eigen solutions of the adjoint  and discrete spectra
\end{center}

Отыщем нули дисперсионного уравнения
$$
\dfrac{\Lambda(z)}{z}=0.
\eqno{(4.1)}
$$
 Для этого разложим  функцию $\lambda_0(z,\alpha)$ в ряд Лорана в окрестности бесконечно
удаленной точки:
$$
\lambda_0(z,\alpha)=-\dfrac{k_2(\alpha)}{z^2}-\dfrac{k_4(\alpha)}{z^4}+\cdots,\qquad
z\to  \infty.
$$
Здесь
$$
k_{2n}(\alpha)=\int_{-\infty}^{\infty}k(\mu',\alpha)\mu'^{2n}d\mu'=
\dfrac{1}{2s_0(\alpha)}\int_{-\infty}^{\infty}f_0(\mu',\alpha)\mu'^{2n}d\mu'=
$$
$$
=\dfrac{s_{2n}(\alpha)}{s_0(\alpha)},\quad
n=0,1,2,\cdots;\quad k_0(\alpha)\equiv 1,
$$
где
$$
s_{2n}(\alpha)=\int_{-\infty}^{\infty}f_0(\mu',\alpha)\mu'^{2n}d\mu'.
$$
Подставим полученное разложение в равенство (3.12). Получаем разложение дисперсионной
функции в ряд Лорана в окрестности бесконечно удаленной точки:
$$
\Lambda(z)=\Lambda_\infty+\dfrac{\Lambda_{-2}}{z^2}+\dfrac{\Lambda_{-4}}{z^4}+\cdots,
\eqno{(4.2)}
$$
где
$$
\Lambda_\infty= 1-\dfrac{1}{w_0}+\dfrac{k_2(\alpha)}{w_0\eta_1^2},\;\qquad
\Lambda_{-2}=\dfrac{k_4(\alpha)-\eta_1^2k_2(\alpha)}{w_0\eta_1^2},
$$
$$
\Lambda_{-4}=\dfrac{k_6(\alpha)-\eta_1^2k_4(\alpha)}{w_0\eta_1^2},\cdots.
$$

Нетрудно видеть, что значение дисперсионной функции в бесконечно удаленной точке
не зависит от химического потенциала и равно
$$
\Lambda_\infty=\Lambda(\infty)=1-\dfrac{1}{w_0}+\dfrac{k_2(\alpha)}{w_0\eta_1^2}=
-\dfrac{\omega^2-\omega_p^2+i \nu\omega}{(\nu-i\omega)^2}.
$$

Отсюда получаем, что
$
\Lambda_\infty \neq 0
$
при любых $\nu\neq 0$, т.е. в любой столкновительной плазме.
Из (4.1) и разложения (4.2) вытекает, что точка $z_j=\infty$ является
нулем дисперсионного уравнения.
Эта точка принадлежит спектру, присоединенному к непрерывному спектру
$(-\infty,+\infty)$. Точке $z_j=\infty$ отвечает следующее решение исходной
системы уравнений (2.8) и (2.10):
$$
H_\infty(x,\mu)=\dfrac{E_\infty}{z_0}\cdot \mu,\qquad e_\infty(x)=E_\infty.
\eqno{(4.3)}
$$

Здесь $E_\infty$ -- произвольная постоянная.

Решение (4.3) не зависит от химического потенциала. Его естественно называть модой Друде.
Оно описывает объемную проводимость плазмы металла, рассмотренную Друде (см., например,
\cite{Ashkroft}).

По определению, дискретным спектром характеристической системы уравнений называется
множество конечных комплексных нулей дисперсионного уравнения,
не лежащих на действительной оси (разрезе дисперсионной функции).

Из разложения (4.2) видно, что в окрестности бесконечно удаленной точки существует
два нуля $\pm \eta_0$ дисперсионной функции $\Lambda(z,\alpha)$:
$$
\pm \eta_0\approx\sqrt{\dfrac{\eta_1^2k_2(\alpha)-k_4(\alpha)}{\eta_1^2(w_0-1)+k_2(\alpha)}}
=\sqrt{\dfrac{\varepsilon(\varepsilon-i\Omega)s_2^2(\alpha)-s_0(\alpha)s_4(\alpha)}
{s_0(\alpha)s_2(\alpha)(1-\Omega^2+i\varepsilon\Omega)}}.
\eqno{(4.4)}
$$

В силу четности дисперсионной функции ее нули различаются лишь знаком.
Под нулем $\eta_0$ будем понимать для определенности такое значение радикала из (4.4), что
$
\Re(w_0/\eta_0)>0.
$
Нулям $\pm\eta_0$, составляющим дискретный спектр характеристической системы,
отвечает следующее решение исходных уравнений
$$
H_{\pm\eta_0}(x_1,\mu)=\exp\Big(-\dfrac{w_0}{\eta_0}x_1\Big)\Phi_1(\eta_0,\mu)+
\exp\Big(\dfrac{w_0}{\eta_0}x_1\Big)\Phi_2(\eta_0,\mu)
\eqno{(4.5)}
$$
и
$$
e_{\pm\eta_0}(x_1)=\Big[\exp\Big(-\dfrac{w_0}{\eta_0}x_1\Big)
+\exp\Big(\dfrac{w_0}{\eta_0}x_1\Big)\Big]E_0.
\eqno{(4.6)}
$$

Здесь $E_0$ -- произвольная постоянная, и
$$
\Phi_1(\eta_0,\mu)=\dfrac{E_0}{w_0}\dfrac{\eta_0\mu-\eta_1^2}{\eta_0-\mu}, \qquad
\Phi_2(\eta_0,\mu)=\dfrac{E_0}{w_0}\dfrac{\eta_0\mu+\eta_1^2}{\eta_0+\mu}.
$$

Решение (4.5) и (4.6) естественно назвать модой Дебая (это -- плазменная
мода). В низкочастотном случае она описывает известное
экранирование Дебая \cite{Ashkroft}.

Из (4.4) видно, что вблизи плазменного резонанса  (при $\Omega\approx 1$), т.е. при
$\omega\approx \omega_p$ модуль нуля $\eta_0$ становится неограниченным,
когда $\varepsilon\ll 1$.

Множество физически значимых параметров $(\Omega,\varepsilon)$ заполняет
четвер\-ть--плоскость
$\{\Omega\geqslant 0, \varepsilon \geqslant 0\}$. Случай $\Omega=0$ (или $\omega=0$) отвечает
внешнему стационарному электрическому полю, а случай $\varepsilon=0$ (или $\nu=0$)
отвечает случаю бесстолкновительной плазмы.

Возникает вопрос, имеются ли еще конечные комплексные нули дисперсионной функции,
помимо нулей $\pm \eta_0$? Точно так же, как и в \cite{LatyshevMono2006}
можно показать, что число
нулей дисперсионной функции равно удвоенному индексу  функции
$G(\tau)=\Lambda^+(\tau)/\Lambda^-(\tau)$ на действительной положительной полуоси,
$
N=2\varkappa(G), \; \varkappa(\alpha)={\rm ind_{[0,+\infty]}}G(\tau).
$
Как и в \cite{LatyshevMono2006} можно показать, что из уравнений
$$
\Re G(\mu,\Omega,\varepsilon,\alpha)=0,\qquad \Im G(\mu,\Omega,\varepsilon,\alpha)=0,\qquad
0\leqslant \mu \leqslant +\infty,
$$
находится кривая $L(\alpha)$, разделяющая области $D^+(\alpha)$ и $D^-(\alpha)$
(см. рис. 3). При
этом  если $(\Omega,\varepsilon)\in D^+(\alpha)$, то число нулей дисперсионной
функции равно двум (это и есть нули $\pm \eta_0$), а если
$(\Omega,\varepsilon)\in D^+(\alpha)$, то дисперсионная функция нулей не имеет.
\begin{figure}[h]
\begin{center}
\includegraphics[width=15.0cm, height=8cm]{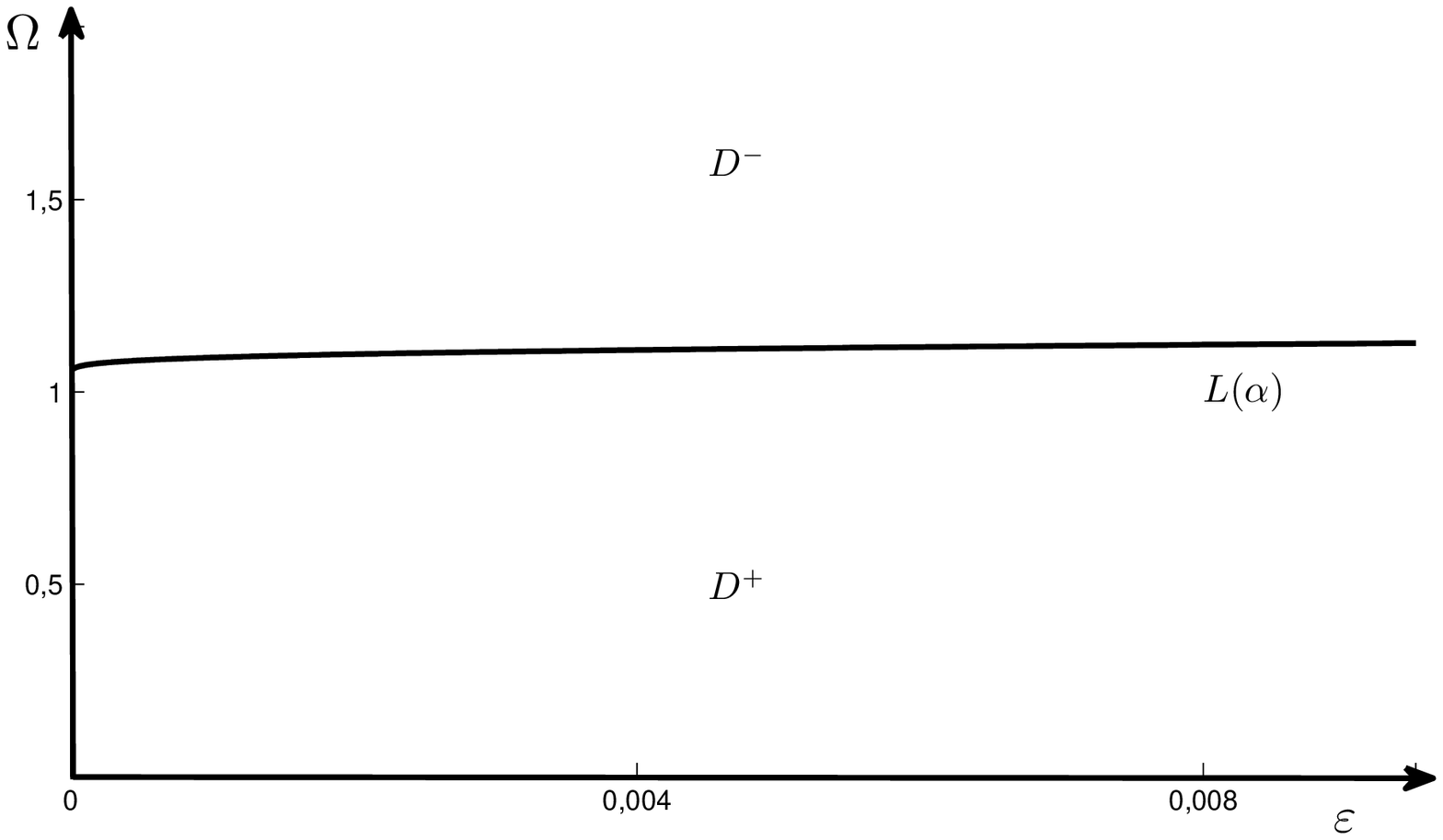}
\end{center}
\begin{center}
{Рис. 3. Случай $\alpha=0$. Кривая $L(\alpha)$ разделяет области единичного $D^+(\alpha)$ и
нулевого $D^-(\alpha)$ индекса. Если $(\Omega,\varepsilon)\in D^+(\alpha)$, то
$N=2$, если $(\Omega,\varepsilon)\in D^-(\alpha)$, то $N=0$.}
\end{center}
\end{figure}

Отметим, что в работе \cite{GranRezhim} разработан метод исследования граничного режима,
когда $(\Omega,\varepsilon)\in L(\alpha)$.

Кривая $L(\alpha)$ определяется параметрическими уравнениями:
$$
L(\alpha):\quad \Omega=\sqrt{L_1(\mu,\alpha)},\quad
\varepsilon=\sqrt{L_2(\mu,\alpha)},\quad 0\leqslant \mu \leqslant +\infty,
$$
где
$$
L_1(\mu,\alpha)=\dfrac{s_0(\alpha)}{s_2(\alpha)}\cdot
\dfrac{\mu^2[\lambda_0(\mu,\alpha)(1+\lambda_0(\mu,\alpha))+s^2(\mu,\alpha)]^2}
{[-\lambda_0(\mu,\alpha)][(1+\lambda_0(\mu,\alpha))^2+s^2(\mu,\alpha)]}
$$
и
$$
L_2(\mu,\alpha)=\dfrac{s_0(\alpha)}{s_2(\alpha)}\cdot\dfrac{\mu^2s^2(\mu,\alpha)}
{[-\lambda_0(\mu,\alpha)][(1+\lambda_0(\mu,\alpha))^2+s^2(\mu,\alpha)]}.
$$ 

\begin{center}
\bf 5. Mirror reflection of electrons from plasma border
\end{center}

Рассмотрим граничную задачу, состоящую из уравнений (2.5) и
(2.6), условия зеркального отражения электронов от границы
плазмы (2.1) и (2.2), и условий (2.3) и (2.4) на электрическое поле.

Покажем, что эта задача  имеет  решение,
представимое в виде разложения по собственным функциям
характеристической системы:
$$
H(x_1,\mu)=\dfrac{E_\infty}{w_0}\mu+\dfrac{E_0}{w_0}
\Bigg[\dfrac{\eta_0\mu-\eta_1^2}{\eta_0-\mu}\exp\Big(-\dfrac{w_0x_1}{\eta_0}\Big)+
\dfrac{\eta_0\mu+\eta_1^2}{\eta_0+\mu}\exp\Big(\dfrac{w_0x_1}{\eta_0}\Big)\Bigg]+
$$
$$
+\int_{-\infty}^{\infty}\Bigg[\exp\Big(-w_0\dfrac{x_1}{\eta}\Big)F_1(\eta,\mu)-
\exp\Big(w_0\dfrac{x_1}{\eta}\Big)F_1(\eta,-\mu)\Bigg]\dfrac{E(\eta)}{w_0}\,d\eta,
\eqno{(5.1)}
$$
$$
e(x_1)=E_\infty+E_0\Big[\exp\Big(-\dfrac{w_0x_1}{\eta_0}\Big)+
\exp\Big(+\dfrac{w_0x_1}{\eta_0}\Big)\Big]+
$$
$$
+\int_{-\infty}^{\infty}\Big[\exp\Big(-\dfrac{w_0x_1}{\eta}\Big)+
\exp\Big(\dfrac{w_0x_1}{\eta}\Big)\Big]E(\eta)\,d\eta.
\eqno{(5.2)}
$$

Здесь $E_0$ и $E_\infty$ -- неизвестные коэффициенты, отвечающие дискретному
спектру ($E_0$ -- амплитуда Дебая, $E_\infty$ -- амплитуда Друде),
$E(\eta)$ -- неизвестная функция, называемая коэффициентом непрерывного спектра.
При $(\Omega,\varepsilon)\in D^-(\alpha)$ в разложениях (5.1) и (5.2) следует
положить $E_0=0$.

Преобразуем разложение (5.1). Для этого воспользуемся нечетностью функции $F_1(\eta,\mu)$
по совокупности переменных: $F_1(-\eta,-\mu)=-F_1(\eta,\mu)$ и четностью функции $E(\eta)$.
Тогда имеем:
$$
\int_{-\infty}^{\infty}\exp\Big(\dfrac{w_0}{\eta}x_1\Big)F_1(\eta,-\mu)E(\eta)\,d\eta
=\int_{-\infty}^{\infty}\exp\Big(-\dfrac{w_0}{\eta}x_1\Big)
F_1(-\eta,-\mu)E(-\eta)\,d\eta=
$$
$$
=-\int_{-\infty}^{\infty}\exp\Big(-\dfrac{w_0}{\eta}x_1\Big)
F_1(\eta,\mu)E(\eta)\,d\eta.
$$

Используя это соотношение, разложения (5.1) и (5.2) представим в следующем виде:
$$
H(x_1,\mu)=\dfrac{E_\infty}{w_0}\mu+\dfrac{E_0}{w_0}
\Bigg[\dfrac{\eta_0\mu-\eta_1^2}{\eta_0-\mu}\exp\Big(-\dfrac{w_0x_1}{\eta_0}\Big)+
\dfrac{\eta_0\mu+\eta_1^2}{\eta_0+\mu}\exp\Big(\dfrac{w_0x_1}{\eta_0}\Big)\Bigg]+
$$
$$
+\dfrac{2}{w_0}\int_{-\infty}^{\infty}
\exp\Big(-\dfrac{w_0x_1}{\eta}\Big)F_1(\eta,\mu)E(\eta)\,d\eta,
$$
$$
e(x_1)=E_\infty+2E_0 \ch\Big(\dfrac{w_0x_1}{\eta_0}\Big)+
2\int_{-\infty}^{\infty}\ch\Big(\dfrac{w_0x_1}{\eta}\Big)E(\eta)\,d\eta.
$$

Начнем со случая $(\Omega,\varepsilon)\in D^+(\alpha)$.
Вычислим значения функции $H(x_1,\pm\mu)$ на верхней границе. Имеем:
$$
H(+l,\pm\mu)=\pm\dfrac{E_\infty}{w_0}\mu+\dfrac{E_0}{w_0}
\Bigg[\dfrac{\pm\eta_0\mu-\eta_1^2}{\eta_0\mp\mu}\exp\Big(-\dfrac{w_0l}{\eta_0}\Big)+
\dfrac{\pm\eta_0\mu+\eta_1^2}{\eta_0\pm\mu}\exp\Big(\dfrac{w_0l}{\eta_0}\Big)\Bigg]+
$$
$$
+\dfrac{2}{w_0}\int_{-\infty}^{\infty}
\exp\Big(-\dfrac{w_0l}{\eta}\Big)F_1(\eta,\pm\mu)E(\eta)\,d\eta.
$$

Подставляя эти равенства в граничное условие (2.2), приходим к интегральному уравнению
Фредгольма второго рода:
$$
\dfrac{E_\infty}{w_0}\mu+\dfrac{E_0\ch({w_0l}/{\eta_0})}{w_0}
\Bigg[\dfrac{\mu\eta_0-\eta_1^2}{\eta_0-\mu}+\dfrac{\mu\eta_0+\eta_1^2}{\eta_0-\mu}\Bigg]+
$$
$$
+\dfrac{1}{w_0}\int_{-\infty}^{\infty}e^{-w_0l/\eta_0}
\Big(F_1(\eta,\mu)-F_1(\eta,-\mu)\Big)E(\eta)\,d\eta=0.
$$

После несложных преобразований приходим к следующему уравнению:
$$
\dfrac{E_\infty}{w_0}\mu+2\dfrac{E_0\ch({w_0l}/{\eta_0})}{w_0}
\mu\dfrac{\eta_0^2-\eta_1^2}{\eta_0^2-\mu^2}+
$$
$$
+\dfrac{2}{w_0}\int_{-\infty}^{\infty}
F_1(\eta,\mu)E(\eta)\ch\dfrac{w_0l}{\eta_0}\,d\eta=0, \quad -\infty<\mu<+\infty.
\eqno{(5.3)}
$$

Можно убедиться непосредственной проверкой, что граничные условия для функции
распределения на нижней пластине (границе) слоя приводят точно к такому же уравнению (5.3).

Подставим разложение (5.2) для электрического поля в граничное условие (5.3).
Получаем следующее уравнение:
$$
E_\infty+2E_0\ch\dfrac{w_0l}{\eta_0}+
2\int_{-\infty}^{\infty}E(\eta)\ch\dfrac{w_0l}{\eta}\,d\eta=1.
\eqno{(5.4)}
$$

Подставим в уравнение (5.3) собственные функции $F_1(\eta,\mu)$. Получаем
сингулярное интегральное уравнение с ядром Коши \cite{Gahov}
на всей числовой оси $-\infty<\mu<+\infty$:
$$
\dfrac{1}{2}E_\infty\mu+E_0\ch\dfrac{w_0l}{\eta_0}\mu\dfrac{\eta_0^2-\eta_1^2}
{\eta_0^2-\mu^2}+\int_{-\infty}^{\infty}
\dfrac{\mu\eta-\eta_1^2}{\eta-\mu}E(\eta)\ch\dfrac{w_0l}{\eta}\,d\eta-
$$
$$
-w_0\eta_1^2\dfrac{\Lambda(\mu)}{\mu k(\mu,\alpha)}E(\mu)\ch\dfrac{w_0l}{\mu}=0.
\eqno{(5.5)}
$$

Введем вспомогательную функцию
$$
M(z)=\int_{-\infty}^{\infty}
\dfrac{z\eta-\eta_1^2}{\eta-z}E(\eta)\ch\dfrac{w_0l}{\eta}\,d\eta,
\eqno{(5.6)}
$$

Функция $M(z)$ аналитична в комплексной плоскости, за исключением разреза --- точек
интегрирования всей числовой прямой $(-\infty,+\infty)$.

Граничные значения вспомогательной функции $M(z)$ сверху и снизу на действительной оси
связаны формулами Сохоцкого:
$$
M^{\pm}(\mu)=\pm \pi i (\mu^2-\eta_1^2)E(\mu)\ch\dfrac{w_0l}{\mu}+
\int_{-\infty}^{\infty}\dfrac{\mu\eta - \eta_1^2}{\eta-\mu}E(\eta)
\ch\dfrac{w_0l}{\eta}\,d\eta,
$$
где интеграл
$$
M(\mu)=\int_{-\infty}^{\infty}\dfrac{\mu\eta-\eta_1^2}
{\eta-\mu}E(\eta)\ch\dfrac{w_0l}{\eta}\,d\eta
$$
понимается как особый в смысле главного значения по Коши.

Из формул Сохоцкого вытекают следующие равенства:
$$
M^+(\mu)-M^-(\mu)=2\pi i(\mu^2-\eta_1^2)E(\mu)\ch\dfrac{w_0l}{\mu},\;\quad
\mu\in(-\infty,+\infty),
\eqno{(5.7)}
$$
$$
M(\mu)=\dfrac{M^+(\mu)+M^-(\mu)}{2},\quad \mu\in(-\infty,+\infty).
$$

С помощью формул Сохоцкого для вспомогательной функции (5.6) и дисперсионной
функции (3.11) преобразуем сингулярное уравнение (5.5) к краевой задаче:
$$
\dfrac{1}{2}E_\infty\mu+
E_0\ch\dfrac{w_0l}{\eta_0}\mu\dfrac{\eta_0^2-\eta_1^2}{\eta_0^2-\mu^2}+
\dfrac{1}{2}\big(M^+(\mu)+M^-(\mu)\big)+
$$
$$
+\dfrac{1}{2}\dfrac{\Lambda^+(\mu)+\Lambda^-(\mu)}{\Lambda^+(\mu)-
\Lambda^-(\mu)}\big(M^+(\mu)-M^-(\mu)\big)=0,\quad -\infty<\mu<+\infty.
$$

Из этого уравнения выводим следующую краевую задачу о нахождении аналитической функции
по ее нулевому скачку на разрезе:
$$
\Lambda^+(\mu)\Bigg[M^+(\mu)+\dfrac{1}{2}E_\infty \mu+E_0\ch\dfrac{w_0l}{\eta_0}\mu
\dfrac{\eta_0^2-\eta_1^2}{\eta_0^2-\mu^2}\Bigg]-
$$
$$
-\Lambda^-(\mu)\Bigg[M^-(\mu)+\dfrac{1}{2}E_\infty \mu+E_0\ch\dfrac{w_0l}{\eta_0}\mu
\dfrac{\eta_0^2-\eta_1^2}{\eta_0^2-\mu^2}\Bigg]=0, \quad
\mu\in(-\infty,+\infty).
\eqno{(5.8)}
$$

Задача (5.8) имеем следующее решение:
$$
M(z)=-\dfrac{1}{2}E_\infty z-\big(E_0\ch\dfrac{w_0l}{\eta_0}\big)
z\dfrac{\eta_0^2-\eta_1^2}{\eta_0^2-z^2}+\dfrac{C_1z}{\Lambda(z)},
\eqno{(5.9)}
$$
где $C_1$ -- произвольная постоянная.

Устраняя полюс в бесконечно удаленной точке, получаем, что
$
C_1=\frac{1}{2}E_\infty \Lambda_\infty.
$
Амплитуда Дебая находится при устранении полюса у решения (5.9)
в точках $\pm \eta_0$. В силу
четности дисперсионной функции эти полюсы устраняются одним условием:
$$
E_0=\dfrac{E_\infty\Lambda_\infty\eta_0}{(\eta_1^2-\eta_0^2)\Lambda'(\eta_0)
\ch({w_0l}/{\eta_0})}.
$$

Коэффициент непрерывного спектра найдем, если подставим решение
(5.9) в формулу Сохоцкого (5.7):
$$
E(\mu)
=\dfrac{E_\infty\Lambda_\infty}{2w_0\eta_1^2}\dfrac{\mu^2\,k(\mu,\alpha)}
{\ch({w_0l}/{\mu})\Lambda^+(\mu)\Lambda^-(\mu)}.
$$

Для нахождения $E_\infty$ воспользуемся уравнением (5.4),
которое перепишем с учетом четности $E(\eta)$:
$$
E_\infty+2E_0\ch\dfrac{w_0l}{\eta_0}+2\int_{-\infty}^{\infty}
E(\eta)\ch\dfrac{w_0l}{\eta}\,d\eta=1,
$$
или в явном виде:
$$
\dfrac{1}{\Lambda_\infty}+\dfrac{2\eta_0}{(\eta_1^2-\eta_0^2)\Lambda'(\eta_0)}+
$$
$$
+\dfrac{1}{2\pi i}
\int_{-\infty}^{\infty}\Big(\dfrac{1}{\Lambda^+(\eta)}-
\dfrac{1}{\Lambda^-(\eta)}\Big)\dfrac{\eta\,d\eta}{\eta^2-\eta_1^2}
=\dfrac{1}{\Lambda_\infty E_\infty}.
\eqno{(5.10)}
$$

Интеграл из (5.10)
$$
J=\dfrac{1}{2\pi i}\int\limits_{-\infty}^{\infty}
\Big(\dfrac{1}{\Lambda^+(\eta)}-\dfrac{1}{\Lambda^-(\eta)}\Big)
\dfrac{\eta\,d\eta}{\eta^2-\eta_1^2}.
$$
можно вычислить аналитически.

Функция
$$
\varphi(z)=\dfrac{z}{\Lambda(z)(z^2-\eta_1^2)},
$$
для которой $\varphi(z)=O(z^{-1})\; (z\to\infty)$, аналитична в верхней и нижней
комплексных полуплоскостях (вне разреза --- действительной оси), за исключением точек
$\pm \eta_1, \pm\eta_0$. Следовательно, этот интеграл равен:
$$
J=\Big[\Res_{\eta_0}+\Res_{-\eta_0}+\Res_{\eta_1}+
\Res_{-\eta_1}+\Res_{\infty}\Big]\varphi(z).
$$
Замечая, что
$$
\Res_{\pm \eta_1}\varphi(z)=\dfrac{1}{2\Lambda_1}, \quad
\Lambda_1=\Lambda(\eta_1)=1-\dfrac{1}{1-i\Omega},\quad
\Res_{\pm \eta_0}\varphi(z)=\dfrac{\eta_0}{\Lambda'(\eta_0)(\eta_0^2-\eta_1^2)},
$$
получаем:
$$
J=\dfrac{2\eta_0}{\Lambda'(\eta_0)(\eta_0^2-\eta_1^2)}+
\dfrac{1}{\Lambda_1}-\dfrac{1}{\Lambda_\infty}.
$$

Подставляя это равенство в (5.10), получаем, что
$$
E_\infty=\dfrac{\Lambda_1}{\Lambda_\infty}
\qquad
C_1=\dfrac{1}{2}\Lambda_1.
\eqno{(5.11)}
$$

Таким образом, разложения (5.1) и (5.2) установлены. Найдены коэффициенты разложений
(5.1) и (5.2). Коэффициент присоединенного спектра $E_\infty$ находится по формуле (5.11),
а коэффициент дискретного спектра $E_0$ и коэффициент непрерывного спектра $E(\eta)$
находятся по формулам:
$$
E_0=\dfrac{\Lambda_1 \eta_0}{(\eta_1^2-\eta_0^2)\Lambda'(\eta_0)\ch({w_0l}/{\eta_0})}
\eqno{(5.12)}
$$
и
$$
E(\eta)=\dfrac{\Lambda_1 \eta_0} {4\pi i (\eta^2-\eta_1^2)\ch({w_0l}/{\eta})}
\Big(\dfrac{1}{\Lambda^+(\eta)}-\dfrac{1}{\Lambda^-(\eta)}\Big).
\eqno{(5.13)}
$$

Структура электрического поля в общем случае такова:
$$
e(x_1)=\dfrac{\Lambda_1}{\Lambda_\infty}+\dfrac{2\Lambda_1\eta_0
}{\Lambda'(\eta_0)(\eta_1^2-\eta_0^2)}\cdot
\dfrac{\ch(w_0x_1/\eta_0)}{\ch(w_0l/\eta_0)}+
$$
$$
+\dfrac{\Lambda_1}{w_0\eta_1^2}\int_{-\infty}^{\infty}
\dfrac{\eta^2 k(\eta,\alpha)}{\Lambda^+(\eta)\Lambda^-(\eta)}\cdot
\dfrac{\ch(w_0x_1/\eta)}{\ch(w_0l/\eta)}\,d\eta.
\eqno{(5.14)}
$$

По формулам (5.11)--(5.13) согласно (5.1) построим и функцию распределения:
$$
H(x_1,\mu)=\dfrac{\Lambda_1}{\Lambda_\infty}\cdot\dfrac{\mu}{w_0}+
\dfrac{\Lambda_1\eta_0}{w_0(\eta_1^2-\eta_0^2)\Lambda'(\eta_0)\ch(w_0l/\eta_0)}\times
$$
$$
\times\Bigg[\exp\Big(-\dfrac{w_0x_1}{\eta_0}\Big)\dfrac{\mu\eta_0-\eta_1^2}{\eta_0-\mu}+
\exp\Big(\dfrac{w_0x_1}{\eta_0}\Big)\dfrac{\mu\eta_0+\eta_1^2}{\eta_0+\mu}\Bigg]+
$$
$$
+\dfrac{\Lambda_1}{w_0^2\eta_1^2}\int_{-\infty}^{\infty}
\dfrac{\exp(-{w_0x_1}/{\eta})F_1(\eta,\mu)\eta^2 k(\eta,\alpha)}
{\Lambda^+(\eta)\Lambda^-(\eta)\ch(w_0l/\eta)}d\eta.
\eqno{(5.15)}
$$

Напомним, что формулы (5.14) и (5.15) справедлива при
$(\Omega,\varepsilon)\in D^+(\alpha)$.
В случае $(\Omega,\varepsilon)\in D^-(\alpha)$ нуль $\eta_0$
дисперсионной функции не существует. Следовательно, можно считать,
что в этом случае $\eta_0=0$; тогда второе слагаемое в
(5.14) и (5.15) пропадает и эти формулы упрощаются.

\begin{center}
  \bf 6. Conclusions
\end{center}

The classical problem about oscillations of the electronic plasmas
in the slab of the conducting medium with arbitrary degree of degeneration
electronic gas, being in the external variable electric field, is analytically solved.
In explicit form the distribution function of electrons and scrinning electric field
in plasma are found. The method applied in work can be applied and to the decision of boundary problems for
systems of Vlasov---Poisson equations (see, for example, \cite {Vedenyapin}).

\end{document}